%% file: paper-v1.tex
\documentclass[conference]{IEEEtran}

\usepackage[pdftex]{graphicx}
\usepackage{textcomp}
\usepackage{pdfcomment}
\usepackage{hyperref}
\usepackage{url}
\usepackage[colorinlistoftodos]{todonotes}

\renewcommand\thesubsection{\arabic{subsection}}

\begin{document}
\title{Challenges and Opportunities in Edge Computing}

\author{\IEEEauthorblockN{Blesson Varghese, Nan Wang, Sakil Barbhuiya, Peter Kilpatrick and Dimitrios S. Nikolopoulos}
\IEEEauthorblockA{School of Electronics, Electrical Engineering and Computer Science\\
Queen's University Belfast, UK\\
Email: \{varghese, nwang03, sbarbhuiya03, p.kilpatrick, d.nikolopoulos\}@qub.ac.uk}
%\and
%\IEEEauthorblockN{Homer Simpson}
%\IEEEauthorblockA{Twentieth Century Fox\\
%\and
%\IEEEauthorblockN{James Kirk\\ and Montgomery Scott}
%\IEEEauthorblockA{Starfleet Academy\\
%San Francisco, California 96678--2391\\
%Fax: (888) 555--1212}
}


%Telephone: (800) 555--1212, Fax: (888) 555--1212}


\maketitle

\begin{abstract}
\input{abstract}
\end{abstract}


\IEEEpeerreviewmaketitle

\section{Introduction}
\label{introduction}
\input{introduction}

\section{Motivation}
\label{motivation}
\input{motivation}

\section{Challenges}
\label{challenges}
\input{challenges}

\section{Opportunities}
\label{opportunities}
\input{opportunities}

\section{Conclusion}
\label{conclusions}
\input{conclusions}

%\section*{Acknowledgment}

%\begin{thebibliography}{00}
%\end{thebibliography}

\bibliographystyle{IEEEtran}  
\bibliography{references}

\end{document}

%% file: abstract.tex
%This position paper argues that there is motivation to look beyond the cloud towards the edge of the network to harness computational capabilities that are currently untapped. This poses at least five research challenges related to general purpose computing on edge nodes, discovering edge nodes, partitioning and offloading tasks, uncompromising quality-of-service and making edge nodes publicly and securely available. Nonetheless, there are five rewarding opportunities in developing standards and benchmarks, frameworks and languages, lightweight libraries and algorithms, micro operating systems and virtualisation and industry-academic collaborations for edge computing.
Many cloud-based applications employ a data centre as a central server to process data that is generated by edge devices, such as smartphones, tablets and wearables. This model places ever increasing demands on communication and computational infrastructure with inevitable adverse effect on Quality-of-Service and Experience. The concept of Edge Computing is predicated on moving some of this computational load towards the edge of the network to harness computational capabilities that are currently untapped in edge nodes, such as base stations, routers and switches. This position paper considers the challenges and opportunities that arise out of this new direction in the computing landscape.

%% file: introduction.tex
The article \textit{`Above the clouds'}\footnote{https://www.eecs.berkeley.edu/Pubs/TechRpts/2009/EECS-2009-28.pdf} presented challenges and opportunities in cloud computing \cite{intro-1}. Cloud research since has rapidly progressed paving the way for many competitors in a crowded marketplace. Although the vision of offering computing as a utility was achieved \cite{intro-2}, the research space is still far from saturation and offers interesting opportunities. 

Many cloud applications are user-driven, which has resulted in opportunities for large-scale data analytics. However, using the cloud as a centralised server simply increases the frequency of communication between user devices, such as smartphones, tablets, wearables and gadgets, we refer to as \textit{edge devices}, and geographically distant \textit{cloud} data centres. This is limiting for applications that require real-time response. Hence, there has been a need for looking \textit{`beyond the clouds'} towards the edge of the network as shown in Figure \ref{figure1}, we refer to as \textbf{\textit{edge computing}} \cite{paper1,paper3}, but is also known as \textbf{\textit{fog computing}} \cite{fogcomputing-1,fogcomputing-2} or \textbf{\textit{cloudlet computing}} \cite{paper2,cloudlet-2}. The aim is to explore possibilities of performing computations on nodes through which network traffic is directed, such as routers, switches and base stations, we refer to as \textit{edge nodes}. The objective of this paper is to define the motivation, challenges and opportunities in edge computing, which is summarised in Figure \ref{figure2}.

%% file: motivation.tex
\begin{figure}
	\centering
    \includegraphics[width=0.5\textwidth]{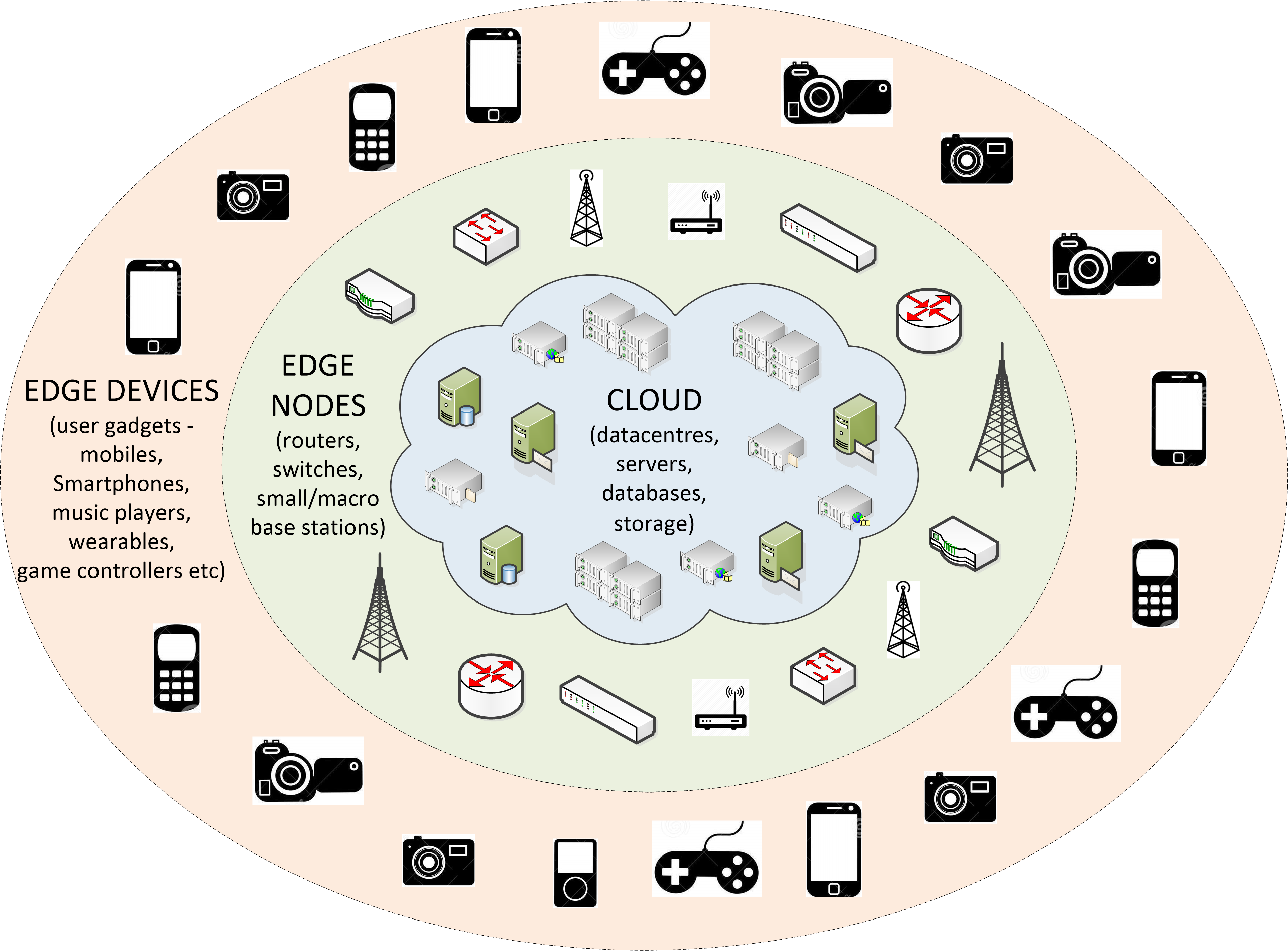}
    \caption{Edge devices and edge nodes in relation to the cloud}
    \label{figure1}
\end{figure}

\begin{figure}
	\centering
    \includegraphics[width=0.5\textwidth]{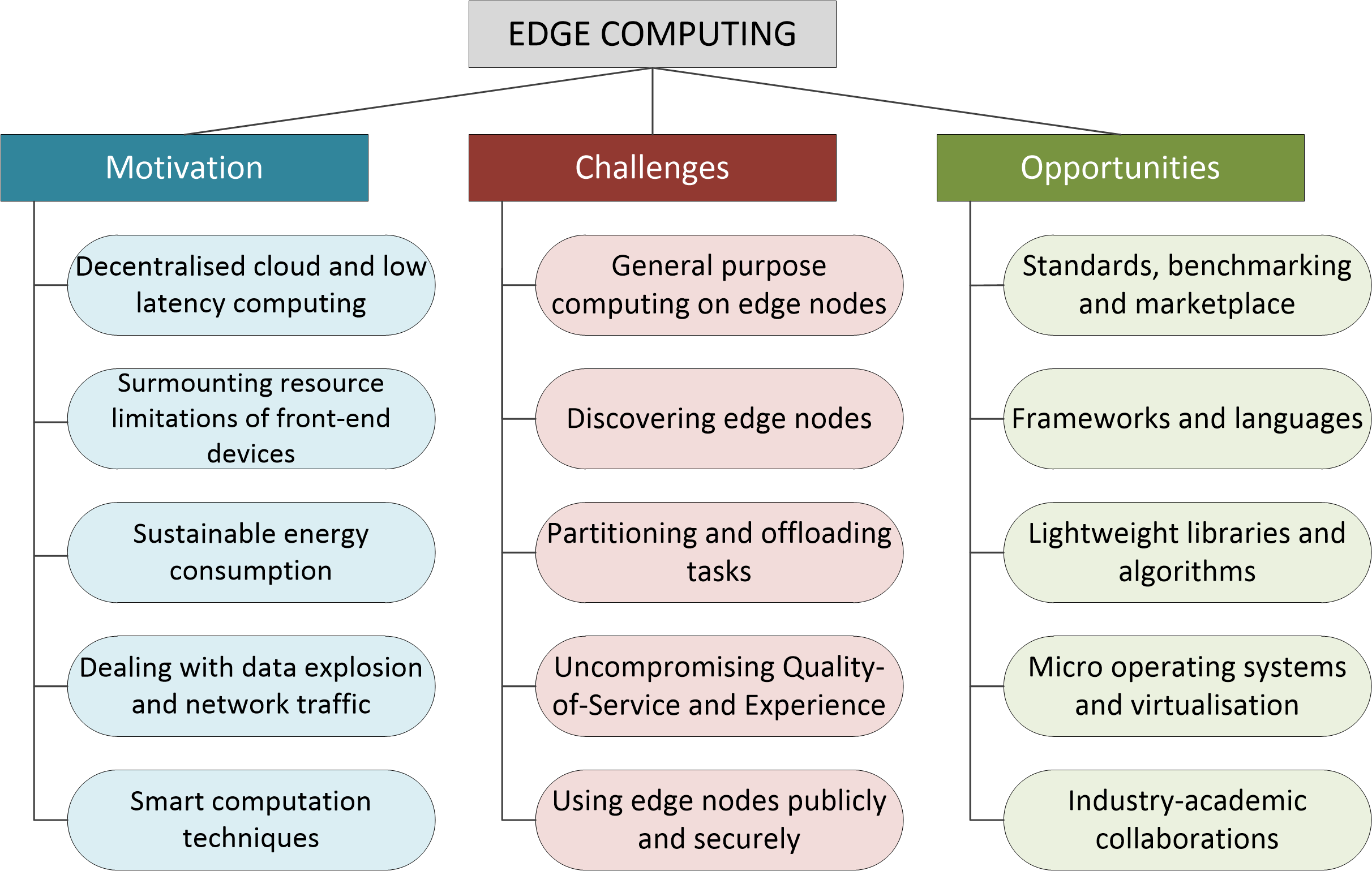}
    \caption{Motivation, challenges and opportunities in edge computing}
    \label{figure2}
\end{figure}

We have identified the following five needs that motivate computing on edge nodes.

\subsection*{1) Decentralised Cloud and Low Latency Computing}
Centralised cloud computing may not always be the best strategy for applications that are geographically distributed. Computing needs to be performed closer to the source of the data to improve the service that is delivered. This benefit can be generalised for any web-based application \cite{zhu2013improving}.
Location-aware applications such as Foursqure\footnote{https://foursquare.com} and Google Now\footnote{https://www.google.co.uk/landing/now/} are becoming popular among mobile users. Computing on edge nodes closer to application users could be exploited as a platform for application providers to improve their service. 

%\subsection{Low Latency Computing}
Numerous data streams are generated by edge devices and making real-time decisions is not possible when analytics is performed on a distant cloud.
Real-time applications, such as a visual guiding service using a wearable camera have a preferred response time between 25ms to 50ms \cite{agarwal2014vision}. The use of current cloud infrastructures poses a serious latency challenge between an edge device and the cloud. The round trip time from Canberra to Berkeley is approximately 175ms \cite{paper2}, which is far from requirements of latency-sensitive applications. Video streaming is the largest mobile traffic generator\footnote{http://www.ericsson.com/res/docs/2016/mobility-report/ericsson-mobility-report-feb-2016-interim.pdf} and satisfying users remains challenging if the bulk of this traffic needs to be channelled to the same source. Similarly, multimedia applications, such as on-demand gaming on current cloud infrastructure, pose similar latency issues for the gamer \cite{choy2012brewing}. Here, to complement the computations performed on the cloud, edge nodes that are located closer (for example, routers or base stations one-hop away from an edge device) to users can be leveraged to reduce network latency. 

\subsection*{2) Surmounting Resource Limitations of Front-end Devices}
User devices, such as smart phones, have relatively restricted hardware resources when compared to a server in a data centre. These front-end devices capture sensory input in the form of text, audio, video, touch or motion and are processed by a service provided by the cloud \cite{CloudBasedAnomalyDetection}. The front-end devices cannot perform complex analytics due to middleware and hardware limitations \cite{CloudMobileAnomalyDetection}; this may sometimes be possible at the expense of draining the battery. Hence, often data needs to be sent to the cloud to meet the computational demands of processing data and meaningful information is then relayed back to the front-end. However, not all data from a front-end device will need to be used by the service to construct analytical workloads on the cloud. Potentially, data can be filtered or even analysed at edge nodes, which may have spare computational resources to accommodate data management tasks.

\subsection*{3) Sustainable Energy Consumption}
There is a significant body of research that has investigated the energy consumption of cloud data centres \cite{datacenterenergy-1, datacenterenergy-2}. Data centres in the next decade are likely to consume three times as much energy consumed today \footnote{http://www.independent.co.uk/environment/global-warming-data-centres-to-consume-three-times-as-much-energy-in-next-decade-experts-warn-a6830086.html} and there is a greater need for adopting energy efficient strategies that can minimise the energy usage\footnote{http://www.computerworld.com/article/2598562/data-center/data-centers-are-the-new-polluters.html}. With more and more applications moving on to the cloud it may become untenable to meet the increasing energy demands. The challenge of increasing energy could be alleviated in a small proportion by incorporating sensible power management strategies in that a number of analytical tasks are performed on edge nodes, such as base stations or routers, closer to the data source instead of overloading data centres with trivial tasks that could perhaps be performed on edge nodes without significant energy implications.  

\subsection*{4) Dealing with Data Explosion and Network Traffic}
The number of edge devices is growing at an enormous rate and one-third of the world's population is expected to have a smart phone by 2018\footnote{http://www.pewglobal.org/2016/02/22/smartphone-ownership-and-internet-usage-continues-to-climb-in-emerging-economies/}\footnote{http://www.telegraph.co.uk/technology/mobile-phones/11287659/Quarter-of-the-world-will-be-using-smartphones-in-2016.html}. Consequently, the volume of data that will be generated will also increase\footnote{http://www.vcloudnews.com/every-day-big-data-statistics-2-5-quintillion-bytes-of-data-created-daily/}; it is anticipated that 43 trillion gigabytes of data will be generated in 2020\footnote{http://www.ibmbigdatahub.com/infographic/four-vs-big-data}. This places the need for expanding data centers to support monitoring and analytical workloads, but this again raises concerns with regard to sustainable energy consumption of data centers. There are attempts to mitigate the energy challenge by performing analytics on the edge device \cite{ids,andromaly,madam}. However, this is restrictive due to resource limitations in the edge devices and collective analytics (of multiple edge devices) cannot be realistically done on an edge device. Another concern with increasing data generation is the volume of network traffic to a central server or cloud thereby reducing the response time of edge devices. This presents the potential of using nodes that are a hop away in the network for complementing computations of the device or of the data center to cope with the growth of data as well as distributing traffic in a network.

%\subsection{Reducing Traffic - Sakil}
%The evolution of smart phone and other edge devices in the recent years has tremendously increased the network traffic to the central server or cloud. This is mainly when the edge devices communicate to the central server for monitoring or feedback purposes. As a result of such increase in the network traffic, monitoring and analysis of edge data in the central server impose a delay in the network response time for the edge devices. Edge computing in the form of pre-processing or filtering of raw data at the edges can help reducing the network traffic, where only the relevant/interesting data are forwarded to the central server.

\subsection*{5) Smart Computation Techniques}
Data generated at a user end needs to be often transported to a cloud server for performing any meaningful analytics which has obvious latency and energy implications. % but may also limit the far-end of the network from being incorporated into critical applications. 
However, there is potential to harness resources at the far end of the network if computation can be hierarchically distributed \cite{distribute-1}. For example, a typical application pipeline may initially filter data generated on the device, after which analytical workloads are executed on the edge nodes through which data is transmitted, before finally arriving at the cloud server where more complex tasks are performed. Alternatively, it may be possible for data centres to offload computations requiring limited resources on to edge nodes, or for the edge nodes to make use of volunteer devices to enhance computational capabilities \cite{offload-1}. Edge nodes can facilitate computations nearer to the source of data (or where data is generated) and can incorporate strategies for remotely enhancing capabilities of front-end devices.

%% file: challenges.tex
Edge computing is still in its infancy and a framework to facilitate this is not yet available. Such frameworks will need to satisfy requirements, such as application development to process requests in real-time on edge nodes. Current cloud computing frameworks, such as the Amazon Web Service\footnote{http://aws.amazon.com}, Microsoft Azure\footnote{https://azure.microsoft.com/en-gb/} and Google App Engine\footnote{https://appengine.google.com/}, can support data-intensive applications, but implementing real-time data processing at the edge of the network is still an open research area \cite{li2015mechanisms,hromic2015real}.
Additionally, the requirement of deploying application workloads on edge nodes will need to be well understood. Deployment strategies - where to place a workload, connection policies - when to use edge nodes and heterogeneity - how to deal with different types of nodes need to be taken into account for deploying applications on the edge. For achieving such a framework, we envisage that the following five research challenges at the hardware, middleware and software layer will need to be addressed. 

\subsection*{Challenge 1 - General Purpose Computing on Edge Nodes}
In theory, edge computing can be facilitated on several nodes that are located between the edge device and the cloud, including access points, base stations, gateways, traffic aggregation points, routers, switches, etc. 
Base stations, for example, incorporate Digital Signal Processors (DSPs) that are customised to the workloads they handle. In practice, base stations may not be suitable for handling analytical workloads simply because DSPs are not designed for general purpose computing. Moreover, it is not readily known if these nodes can perform computations in addition to their existing workloads. The OCTEON Fusion\textregistered
Family\footnote{http://www.cavium.com/OCTEON-Fusion.html} by CAVIUM, a small cell "Base Station-on-a-chip" family, scales from 6 to 14 cores to support users ranging from 32 to 300+. Such base stations could perhaps be used during off-peak hours to exploit the computational capabilities of multiple computing cores available.  
A number of commercial vendors have taken a first step to realise edge computing using software solutions. 
For example, Nokia's software solution\footnote{http://networks.nokia.com/portfolio/solutions/mobile-edge-computing\#tab-highlights} for mobile edge computing (MEC) aims to enable base station sites for edge computing. Similarly, Cisco's IOx\footnote{http://www.cisco.com/c/en/us/products/cloud-systems-management/iox/index.html} offers an execution environment for its integrated service routers. These solutions are specific to hardware and hence may not be suitable in a heterogeneous environment. One challenge in the software space will be to develop solutions that are portable across different environments.  

There is research in upgrading the resources of edge nodes to support general purpose computing. For example, a wireless home router can be upgraded to support additional workloads \cite{meurisch2015upgrading}. 
Intel's Smart Cell Platform\footnote{http://www.intel.com/content/www/us/en/communications/smart-cells-revolutionize-service-delivery.html} uses virtualisation for supporting additional workloads. Replacing specialised DSPs with comparable general purpose CPUs gives an alternative solution but this requires a huge investment. 

\subsection*{Challenge 2 - Discovering Edge Nodes}
Discovering resources and services in a distributed computing environment is an area that is well explored. This is facilitated in both tightly and loosely coupled environments through a variety of techniques that are incorporated into monitoring tools \cite{monitoring-0,monitoring-1,monitoring-2,monitoring-3} and service brokerages \cite{broker-1, broker-2, broker-3, broker-4}. Techniques such as benchmarking underpin decision-making for mapping tasks onto the most suitable resources for improving performance. 

However, exploiting the edge of the network requires discovery mechanisms to find appropriate nodes that can be leveraged in a decentralised cloud set up. These mechanisms cannot be simply manual due to the sheer volume of devices that will be available at this layer. Moreover, they will need to cater for heterogeneous devices from multiple generations as well as modern workloads, for example large scale machine learning tasks, which were previously not considered. Benchmarking methods will need to be significantly rapid in making known the availability and capability of resources. These mechanisms must allow for seamless integration (and removal) of nodes in the computational workflow at different hierarchical levels without increasing latencies or compromising the user experience. Reliably and proactiviely dealing with faults on the node and autonomically recovering from them will be desirable. Existing methods used in the cloud will not be practical in this context for the discovery of edge nodes. 

%\subsection{Transforming the Edge Nodes to General Purpose Computing Nodes - Nan}
%How can general purpose computing be done on these devices
%Currently edge-node-to-be were designed specifically to their basic services and it involves non-trivial tasks to transform or upgrade them for the realisation of edge computing. Firstly, additional hardware needs to be identified. \cite{meurisch2015upgrading} is among the early initiatives to upgrade wireless home routers to enable edge computing. This work provides some insight into edge node transformation but for different edge nodes, the optimal choice on processor, memory, storage unit will be different and the original specifications of these edge node devices such as physical size and compatibility need to be considered. Intel's Smart Cell Platform\footnote{http://www.intel.com/content/www/us/en/communications/smart-cells-revolutionize-service-delivery.html} tries to transforming base station through virtualisation. Replacing the specialised DSPs with general-purpose CPUs gives an alternative solution but this will introduce additional capital expenditure for base station owners to abandon their existing equipment. Secondly, it is important to form collaborations between hardware designers of these edge nodes and developers for edge computing platforms and applications. Some of the edge nodes such as mobile base stations have complex hardware designs thus a good understanding of what they can do beyond their basic services is a prerequisite for further transformations.

\subsection*{Challenge 3 - Partitioning and Offloading Tasks}
Evolving distributed computing environments have resulted in the development of numerous techniques to facilitate partitioning of tasks that can be executed at multiple geographic locations \cite{partitioning-1, partitioning-2}. For example, workflows are partitioned for execution in different locations \cite{wflow-1,wflow-2}. Task partitioning is usually expressed explicitly in a language or management tool.  

However, making use of edge nodes for offloading computations poses the challenge of not simply partitioning computational tasks efficiently, but doing this in an automated manner without necessarily requiring to explicitly define the capabilities or location of edge nodes. The user of a language that can leverage edge nodes may anticipate flexibility to define a computation pipeline - hierarchically in sequence (first at the data centre then at the edge nodes or first at the edge node and then at the data centre) or potentially over multiple edge nodes simultaneously. Inherently, there arises the need for developing schedulers that deploy partitioned tasks onto edge nodes.   

%\subsection{Uncompromising User Experience - Sakil}
%Computations away from the device, What should be done to not affect users.
%Cloud computing brings the benefits for both the Cloud users and the Cloud service providers. Similarly, we expect Edge computing to benefit both the edge device users and edge service providers. However, in the case of Edge computing, the user-provider scenario is bit different as Edge computing utilises user-end devices/nodes for the computation, which may introduce some overhead on the usual services that the edge device/node carries out. This overhead may disrupt the edge device services when we consider the devices with constrained resources. Therefore, estimating carefully on the overhead of running Edge computation on the edge devices should be the preliminary step for any Edge computing implementation. In any case, the user-experience due to the implementation of Edge computation should not be compromised.  

\subsection*{Challenge 4 - Uncompromising Quality-of-Service (QoS) and Experience (QoE)}
Quality delivered by the edge nodes can be captured by QoS and quality delivered to the user by QoE. One principle that will need to be adopted in edge computing is to not overload nodes with computationally intensive workloads \cite{beck2014mobile,simoens2015challenges}.
%\pdfcomment{Add text: , which is a critical requirement for several works done on edge computing \cite{beck2014mobile,simoens2015challenges}}.
The challenge here is to ensure that the nodes achieve high throughput and are reliable when delivering for their intended workloads if they accomodate additional workloads from a data center or from edge devices. Regardless of whether an edge node is exploited, the user of an edge device or a data centre expects a minimum level of service. For example, when a base station is overloaded, it may affect the service provided to the edge devices that are connected to the base station. A thorough knowledge of the peak hours of usage of edge nodes is required so that tasks can be partitioned and scheduled in a flexible manner. The role of a management framework will be desirable but raises issues related to monitoring, scheduling and re-scheduling at the infrastructure, platform and application levels\footnote{https://portal.etsi.org/portals/0/tbpages/mec/docs/mobile-edge\_computing\_-\_introductory\_technical\_white\_paper\_v1\%2018-09-14.pdf}. 

\subsection*{Challenge 5 - Using Edge Nodes Publicly and Securely}
Hardware resources that are owned by data centres, supercomputing centres and private organisations using virtualisation can be transformed to offer computing as a utility. The associated risks for a provider and users are articulated\footnote{http://www.cloud-council.org/deliverables/CSCC-Practical-Guide-to-Cloud-Service-Agreements.pdf}, thereby offering computing on a pay-as-you-go basis. This has resulted in a competitive marketplace with numerous options and choices to satisfy computing consumers by meeting Service Level Agreements (SLAs) \cite{sla-1}.

However, if alternative devices, such as switches, routers and base stations, need to be used as publicly accessible edge nodes a number of challenges will need to be addressed. Firstly, the risk associated by public and private organisations that own these devices as well as those that will employ these devices will need to be articulated. Secondly, the intended purpose of the device, for example, a router managing internet traffic, cannot be compromised when used as an edge computing node. Thirdly, multi-tenancy on edge nodes will only be possible with technology that places security as a prime concern. Containers, for example, a potential lightweight technology usable on edge nodes will need to demonstrate more robust security features \cite{containersecurity-1}. Fourthly, a minimum level of service will need to be guaranteed to a user of the edge node. Fifthly, the workloads, computation, data location and transfer, cost of maintenance and energy bills will need to be considered for developing suitable pricing models to make edge nodes accessible.

%% file: opportunities.tex
Despite challenges that arise when realising edge computing, there are numerous opportunities for academic research. We identify five such opportunities. 

\subsection*{Opportunity 1 - Standards, Benchmarks and Marketplace}
Edge computing can be realised in practice and be made accessible publicly if responsibilities, relationships and risks of all parties involved are articulated. There are numerous efforts to define a variety of cloud standards, such as by National Institutes of Standards and Technology(NIST)\footnote{http://www.nist.gov/itl/cloud/}\footnote{http://nvlpubs.nist.gov/nistpubs/Legacy/SP/nistspecialpublication800-145.pdf}, IEEE Standards Association\footnote{http://cloudcomputing.ieee.org/standards}, International Standards Organisation (ISO)\footnote{http://www.iso.org/iso/home/store/catalogue\_tc/catalogue\_tc\_browse.htm?\\commid=601355}, Cloud Standards Customer Council (CSCC)\footnote{http://www.cloud-council.org/} and the International Telecommunication Union (ITU)\footnote{http://www.itu.int/en/ITU-T/jca/Cloud/Pages/default.aspx}. However, such standards will now need to be reconsidered in light of additional stakeholders, such as public and private organisations that own edge nodes, to define the social, legal and ethical aspects of using edge nodes. This is certainly not an easy task and requires commitment and investment from public and private organisations and academic institutions. 

Standards can be implemented only if the performance of edge nodes can be reliably benchmarked against well known metrics. Benchmarking initiatives for the cloud include those by the Standard Performance Evaluation Corporation (SPEC)\footnote{https://www.spec.org/benchmarks.html\#cloud} and by numerous academic researchers \cite{benchmarking-1,benchmarking-3,benchmarking-2,benchmarking-4}. In a noisy environment, such as the cloud, benchmarking poses significant challenges. The current state-of-the-art is not yet mature and significant research is required to deliver comprehensive benchmarking suites that can gather metrics accurately. Benchmarking edge nodes will therefore be more challenging, but opens new avenues of research. 

Using edge nodes is an attractive prospective when responsibilities, relationships and risks are defined. Similar to a cloud marketplace, an edge computing marketplace that offers a variety of edge nodes on a pay-as-you-go basis is feasible. Research in defining SLAs for edge nodes and pricing models will be required to create such a marketplace.

\subsection*{Opportunity 2 - Frameworks and Languages}
There are many options to execute applications in the cloud paradigm. In addition to popular programming languages, there is a wide variety of services to deploy applications on the cloud. When resources outside the cloud are employed, for example, running a bioinformatics workload on the public cloud where input data is obtained from a private database, a workflow is usually employed. Software frameworks and toolkits for programming large workflows in a distributed environment is a well defined research avenue \cite{pegasus-1,taverna-1}.

However, with the addition of edge nodes that may support general purpose computing, there will be the need for developing frameworks and toolkits. The use cases for edge analytics are likely to differ from existing workflows, which are mostly explored for scientific domains, such as bioinformatics \cite{biowf-1} or astronomy \cite{astrowf-1}. Given that edge analytics will find its use-cases in user-driven applications, the existing frameworks may not be appropriate to express an edge analytics workflow. The programming model that aims to exploit edge nodes will need to support task and data level parallelism and at the same time execute workloads on multiple hierarchical levels of hardware. The language that supports the programming model will need to take into account the heterogeneity of hardware and the capacity of resources in the workflow. If edge nodes are more vendor specific, then the frameworks supporting the workflow will need to account for it. This is more complex than existing models that make the cloud accessible.

\subsection*{Opportunity 3 - Lightweight Libraries and Algorithms}
Unlike large servers edge nodes will not support heavyweight software due to hardware constraints. For instance, a small cell base station with Intel's T3K Concurrent Dual-Mode system-on-chip (SoC)\footnote{http://www.intel.com/content/dam/www/public/us/en/documents/solution-briefs/transcede-t3k-solution-brief.pdf} has a 4-core ARM-based CPU
and limited memory, which will not be sufficient for executing complex data processing tools such as Apache Spark\footnote{http://spark.apache.org} that requires at least 8 cores CPU and 8 gigabyte memory for good performance. Edge analytics require lightweight algorithms that can do reasonable machine learning or data processing tasks \cite{kartakis2014real, santos2013dial}. Apache Quarks\footnote{http://quarks.incubator.apache.org}, for example, is a lightweight library that can be employed on small footprint edge devices such as smart phones to enable real-time data analytics. However, Quarks supports basic data processing, such as filtering and windowed aggregates, which are not sufficient for advanced analytical tasks such as context-aware recommendations. Machine learning libraries that consume less memory and disk usage would benefit data analytical tools for edge nodes. TensorFlow\footnote{https://www.tensorflow.org/}
%\cite{2016arXiv160304467A}
 is another example framework that supports deep learning algorithms and supports heterogeneous distributed systems, but its potential for edge analytics is still to be explored. 

\subsection*{Opportunity 4 - Micro Operating Systems and Virtualisation}
Research in micro operating systems or microkernels can provide inroads to tackling challenges related to deployment of applications on heterogeneous edge nodes. Given that these nodes do not have substantial resources like in a server, the general purpose computing environment that is facilitated on the edge will need to exhaust fewer resources. The benefits of quick deployment, reduced boot up times and resource isolation are desirable \cite{MobileVirtualisation}. There is preliminary research suggesting that mobile containers that multiplex device hardware across multiple virtual devices can provide similar performance to native hardware \cite{Andrus:2011}. Container technologies, such as Docker\footnote{https://www.docker.com} are maturing and enable quick deployment of applications on heterogeneous platform. More research is required to adopt containers as a suitable mechanism for deploying applications on edge nodes. 

%One of the major challenges in Edge computing is the resource availability for the computation. This is because most of the Edge nodes or devices do not have general purpose computing infrastructure. Micro operating systems or microkernels can provide optimal solutions for deploying Edge applications into the heterogeneous edge nodes or devices, which are constrained by hardware resources. In particular, edge computing can leverage the benefits of micro operating systems such as quick deployment, reduced boot up times, and low resource requirements. Furthermore, mobile virtualisation based on microkernel architecture can enhance security and improve resource usage due to the isolation and load-balancing feature offered by the virtualisation technology. Andrus et al. \cite{Andrus:2011} proposed a mobile container (OS-level virtualisation) based on Android system, which multiplexes device hardware across multiple virtual devices and provides native hardware device performance. In addition, application-based containers like Docker containers  can help to ship and run an Edge application on any heterogeneous edge platform.

\subsection*{Opportunity 5 - Industry--Academic Collaborations}
Edge computing offers a unique opportunity for academia to re-focus its research activities broadly in applied distributed computing, specifically within cloud and mobile computing. It is not easy for academic research to focus on scale without making assumptions that may not always correspond to reality. This is simply because a large number of academic institutions and researchers cannot gain access to infrastructure owned by the industry or government to validate and refine their research \cite{academiccloudcomp-2}. Premier academic institutions that have reliable industry and government relations have however produced more meaningful and impactful research~\cite{academiccloudcomp-1}. 

Research in the edge computing space can be driven by an open consortium of industry partners, such as mobile operators and developers, gadget developers and cloud providers, as well as interested academic partners to the mutual benefit of both.

%% file: conclusions.tex
Edge computing is still in its infancy and has potential to pave the way for more efficient distributed computing. This paper has highlighted the significant potential of computing at the edge of the network and presented five research challenges and five rewarding opportunities in realising edge computing.